# Four checks for low-fidelity synthetic data: recommendations for disclosure control and quality evaluation

Gillian M Raab,  Scottish Centre for Administrative Data Research    gillian.raab@ed.ac.uk

Sophie McCall. Research Data Scotland sophie.mccall@researchdata.scot

Liam Cavin,  National Records of Scotland ,  liam.cavin@nrscotland.gov.uk

## Abstract

Confidential administrative data is usually only available to researchers within a trusted research environment (TRE). Recently, some UK groups have proposed that low-fidelity synthetic data (LFSD) is available to researchers outside the TRE to allow code-testing and data discovery. There is a need for transparency so that those who access LFSD know how it has been created and what to expect from it.

Relationships between variables are not maintained in LFSD, but a real or apparent **data breach** can occur from its release. To be useful to researchers for preliminary analyses LFSD needs to meet some minimum quality standards. Researchers who will use the LFSD need to have details of how it compares with the data they will access in the TRE clearly explained and documented.

We propose that these checks should be run by data controllers before releasing LFSD to ensure it is well documented, useful and non-disclosive.

1. **Labelling**
   To avoid an apparent data breach, steps must be taken to ensure that the SD is clearly identified as not being real data.
2. **Disclosure**
   The LFSD should undergo disclosure risk evaluation as described below and any risks identified mitigated.
3. **Structure**
   The structure of the SD should be as similar as possible to the TRE data.
4. **Documentation**
   Differences in the structure of the SD compared to data in the TRE must be documented, and the way(s) that analyses of the SD expect to differ from those of data in the TRE must be clarified.

We propose details of each of these below; but a strict, rule-based approach should not be used. Instead, the data holders should modify the rules to take account of the type of information that may be disclosed and the circumstances of the data release (to whom and under what conditions).

# 1. Introduction and background

In the UK and elsewhere there is increased interest in using administrative data for research to inform policy in area such as health, education and social policy [1,2] . Several UK organisations have been established to facilitate making data from administrative sources available to researchers. They also support the linkage of such administrative data to other sources, such as Censuses or government surveys. The Economic and Social Research Council funds ADR UK (Administrative Data Research UK)[1] and its partner organisations in England, Scotland, Wales and Northern Ireland[2] ,to work with Government departments and National Statistics agencies to make administrative data available to researchers. HDR UK[3] performs a similar function for Health data and Research Data Scotland[4] integrates data from many sources in Scotland. Researchers must apply for access to the data that is usually available in a trusted research environment (TRE). The process of applying and gaining access to the data in the TRE can be lengthy. The provision of synthetic data to researchers can allow them to develop their analyses and thus shorten the time between completing an application and obtaining results. The benefit of making synthetic data available are even greater where a visit to a safe setting is required to access the TRE. This is the case for the Scottish Longitudinal Study that provides a synthetic data option for users [3,4]. Van Kestern [5] has argued that synthetic data are required to democratise research with sensitive data.

Methods for creating synthetic data (SD) for disclosure control have developed over the 30 years since this was first proposed [6] as described in two recent reviews [7,8]. Synthetic data sets can be created using information from the original data, usually held within the TRE, for several purposes.

    **Purpose 1**: to allow users to discover the features of the data in the TRE and to develop code that will then be run on the data available in the TRE.

    **Purpose 2:** to allow users to develop plans of analysis by using relationships in the synthetic data that will be expected to approximate those in found from the TRE data

    **Purpose 3:** to use in training researchers to understand the methodologies they will need to use data sets that are only available in TREs [9].

Kokosi et al [10] provide an overview of synthetic administrative data for research based on a categorisation suggested by the UL Office of National Statistics (ONS) [11]. They categorise synthetic data on a spectrum from *low fidelity* (Minimally disclosive, minimal analytic value) to *high fidelity* (more disclosive, more analytic value)[5]. **Purpose 1** can be achieved by SD at the low end of the spectrum. **Purposes 2** requires the highest-possible fidelity, since preliminary analyses can influence the final results of an investigation. **Purpose 3** requires at least moderate fidelity to ensure the training data appear realistic.

---

[1] Administrative Data Research UK https://www.adruk.org/
[2] https://www.adruk.org/about-us/our-partnership/
[3] Health Data Research UK https://www.hdruk.ac.uk/
[4] Research Data Scotland  https://www.researchdata.scot/

[5] See Table 1 in the paper and an extended version in the supplementary material

It is difficult to produce high fidelity SD that will reproduce all the complex relationships between variables in large administrative data bases. There is also concern about possible disclosure risk from SD that matches the original data too well. Although there has been some recent work on methods of assessing disclosure risk from SD [12.13], experience with these metrics is limited. Thus several UK organisations have suggested that the priority for holders of sensitive administrative data should be to produce LFSD for **Purpose 1** [14, 15,16] . Pilot projects have been evaluated (e.g. [15]) and at least one example of LFSD is now available [17]

Even LFSD needs to be evaluated for usefulness and disclosure risk. The goal of this short paper is to outline the steps that a data custodian should take to ensure that their LFSD is useful and does not pose a potential disclosure risk.

## 2 Steps in preparing synthetic data

Figure 1 illustrates the steps used in making data available for researchers. A **raw data set** is produced at step 1 either by downloading from an administrative data system or from survey or census data. The raw data may be from a single source, or from linkage of different sources.

At step 2 the data is prepared to make it suitable for analysis to make it suitable for analysis. This may involve steps to reduce disclosure risk (anonymisation and suppression of potentially disclosive values) and to improve utility (correction of inconsistent values and imputation of missing values or missing records). At this step a **research-ready** data set is created. A review by Garth-Lone et al.[18] identifies the characteristics that make administrative data research-ready. They emphasise that the data needs to be curated to ensure its quality and documented to ensure transparency. Research-ready data may be used for internal analyses or supplied to external researchers either as an open public-use file (PUF) file[6], or for restricted use by researchers to analyse on their own machines or to access in a TRE. Steps to limit disclosure control at step 2 will depend crucially on the setting in which the data will be released [19]. These include statistical disclosure control (SDC) procedures such as rounding and top- or bottom-coding for numerical variables, aggregation of categories and record swapping [20,21].

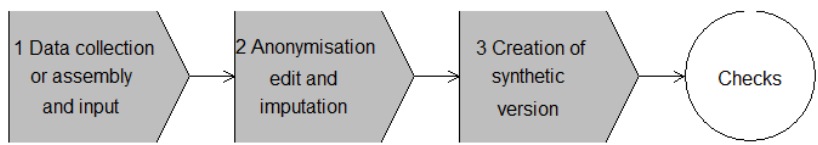

*Figure 1  Steps in preparing data for synthesis.*

At the final step the data in the TRE is used to create synthetic data sets. SDC procedures can also be applied to SD before it is released to researchers. Before its release SD should undergo a series of checks. Those we propose here are appropriate for LFSD created by methods described in Section 3.

---
[6] Another way that open data is supplied is the data set that sits behind a flexible table builder that allows external users to create their own tables (Thompson et al. 2013).

## 3 Methods for creating LFSD

Different methods can be employed at the low end of the SD spectrum. Here we will consider LFSD methods that do not attempt to maintain the relationships between variables. They are usually created by one of two methods:

1. By creating columns of data that match the structure and range of values in the original, but using only the metadata supplied for data in the TRE; here referred to as "**from metadata**"**.** This method does not require access to the data in the TRE. It is sometimes referred to as "Dummy data".
2. By taking independent samples from the univariate distributions of each column in the original data; here referred to as "**from margins**".

The samples size of the SD need not be the same as then original. Small sample sizes for SD can reduce disclosure risk (see below) and make it clear that it is not the original. Even within these relatively simple methods the usefulness of the SD can be enhanced by anticipating how the data will be used. For example, if medical data consist of dates of diagnosis and dates of death, then people in LFSD might be diagnosed after they die. To avoid this the data should be transformed into date of diagnosis followed by length of survival. Another example occurred in [15] where the data set had a series of columns for components of income, as well as a column for total income.

## 4 Disclosure risk (DR)

Data controllers have a responsibility to the people or other units whose data they hold. This includes ensuring that information about an individual unit is not made available to people who are not entitled to know it. Such an incident is referred to as a **data breach**. A data breach can have adverse consequences for the unit and may lead to loss of reputation and of trust for the organisation that holds the data. A data breach requires:

- information about a unit appearing to be learned from the data and
- information to be disseminated to others who do not already know it are not entitled to do so.

Such information may or may not be true. When it is false the consequences for the unit and the organisation may be as serious as when it is true. Hence the importance of making it clear to the person viewing the data that it is not the original.

Techniques for ensuring that the user knows that the data being viewed are synthetic include appropriate labelling and are detailed as the first check in Section 6. In addition, a synthetic data set that is much smaller than the original will make the user realise that only a subset of the data have been synthesized so that records with unique key combinations in the subsample are unlikely to identify those unique in the original.

There are two types of DR:

- Identity disclosure risk (IDR)– learning that an identified unit is in the data
- Attribute disclosure – learning something new about an identified unit

Each depend on what **keys** can be assumed to be known for subjects in the data. For people these could be (e.g. age, sex, postcode, education) or for businesses (e.g. year founded, number of employees, location). The term quasi-identifiers is often used for such information. Attribute disclosure for SD depends on being able to identify a record, and then learn the value of other items, so controlling identity disclosure should be an adequate precaution for LFSD,

There may still be a real identity DR for LFSD, ***even when someone knows that it is synthetic***. If there is a unique value in one column of the data (e.g. exact value of a person's income or the exact date a business was founded) then someone with knowledge of that value will assume that the subject is there. These risks can be mitigated by the following steps that may often have been carried out when research-ready data are prepared (step 2 in Figure 1).

- giving data with limited precision by e.g. giving income in units of £1000 or year only for dates,
- disguising any extreme values by top or bottom coding values at the extremes of the range (e.g. outside the range of the 1st and 99$^{th}$ percentile), or by excluding values at the extremes of the range where the number of records outside the range is below a threshold[7].
- pooling or supressing categories with small numbers of individuals in the original (e.g. county of birth with fewer than 5 individuals).

SD may appear to pose an IDR if it includes records that have unique combinations of these keys. These records should then be checked to find if their combination of values exists in the original data and whether it is unique in the original[8]. Those that are unique in both the SD and the original are designated as *replicated uniques* and those unique in the SD and present in the original as *uniques in the original*[9]. Where these categories are a small proportion of all records they can be excluded from the SD. Nowok [22] found that the removal of *replicated uniques* from high fidelity synthetic data had very little impact on utility. If there are many records in these categories, then one or more of the keys needs to be modified to reduce this number. For example, when one of the keys is a geographic area for the unit, it can be replaced by one for a larger area (e.g. give Local Authority area rather than postcodes as the geographic key).

Our investigations show that LFSD often has a higher proportion of apparently disclosive records than the original, and as one would expect, this is especially true when the sample size is smaller than the original. But a much higher proportion of these are not present in the original, so that rates of *replicated uniques or uniques in the original* are low, and in our experience lower than for high-fidelity synthesis of the same original.

---

[7] The second of these is more appropriate, especially for bottom coding.
[8] An alternative criterion would be to exclude records with counts of (say) 3 or fewer in the synthetic data
[9] The second category will include more records than the first, but may be more appropriate. It is based on a scenario of a person who believes the SD to be real and thinks they recognise a unit in the SD. Such a person would not know whether the apparent matching record is unique in the original. Thus the wider category may be more appropriate.

# 5 Utility for low-fidelity synthetic data

Although the relationship between variables in the original is not maintained in the SD, there are some aspects of the SD that need to be maintained to make it useful for developing code:

- The variables in the SD must be a subset of those in the original with the same names (allowing for the addition of any prefixes or suffices added to the SD to fulfil (see Check 1c in section 6).
- The levels/value labels of categorical variables must be identical except for any pooled categories in the SD.
- Whether a variable has any missing values must agree between the SD and the original.
- The precision to which any numeric data are given (e.g. Number of decimal places quoted), should be the same in the SD and the original.

The metadata for the SD will usually point to the metadata for the original as the source of details, but it should also contain details of where the two differ. It should also explain which relationships in the original expect to be maintained in the SD. This will usually be

- For LFSD from metadata – no tables calculated from the SD will be similar to the same tables from the original.
- For LFSD from margins – only tables from the SD for one variable at a time will be similar to those from the original. No other tables or regression model results will be expected to be similar for the SD compared to the original.

# 6 Four checks for data controllers

Here we summarise a possible protocol that might be followed to ensure LFSD will be useful for researchers and avoid any real or apparent data breaches.

1. **Label**
   a. Make it clear in the header of the metadata for the SD that this is NOT the original
   b. Ensure that the filename for the SD includes the word 'synthetic'
   c. Give variable names in the SD a prefix such as 'synth_' or a suffix such as '_synth'.

2. **Check and reduce disclosure**
   a. Identify any variables in the original where a unique value would identify an individual, such as exact values of numeric variables or rare categories for grouped variables.

b. Define **keys**, as described in Section 5, and check the SD for the proportion of records in the SD that are *replicated uniques* or *uniques in the original* (see above).
   c. If necessary modify the SD with one or more of the following
      i. reduce the detail provided in the key variables to reduce the proportion of *replicated uniques* or *uniques in the original* to a small proportion
      ii. reduce the precision of numeric data in the SD.
      iii. pool rare categories for categorical data.
      iv. remove records that are *replicated uniques* or *uniques in the original*

3. **Maintain structure**
   a. Ensure that variable names in the SD are the same as in the TRE data, allowing for 1c.
   b. Set group names for categorical data variables in the SD as exact matches to those in the data in the TRE data, allowing for 2ciii.
   c. Check that the presence of missing values in variables agrees between original and SD

4. **Document differences**
   a. create metadata for the SD that points to the metadata for the TRE data and document any differences between the SD and the TRE data.
   b. Include a clear statement in the metadata describing the relationships that will expect to be maintained in the SD compared to the TRE data.

# 7 Conclusions

Our first proposed check (labelling) is probably the most important. Researchers should adhere to whatever handling instructions are provided alongside the data, such as any restrictions on sharing it. However, careful labelling of SD by the creator can mitigate the ensuing risks, even if handling instructions are not followed and wider sharing does occur. Labelling is even more critical where SD are freely available without the need to register and agree to sharing restrictions.

While LFSD will be expected to have a low disclosure risk, this risk may not be zero even for LFSD created from metadata. For example, metadata may give the range of actual values for numeric variables and the highest of lowest value may identify a small number of individuals, or even be unique in the original data. A similar situation can occur for an infrequent code for a categorical variable. Our proposals on checking for *replicated uniques* or *uniques in the original* are based on recent work for fully SD [23]. Some people may argue that they are not necessary for LFSD, but we think that some attempt should be made to measure these risks, especially for data that is relatively freely available.

Steps 3 and 4 are important to make the LFSD useful to researchers. There are other aspects of usefulness that we have not specified as a necessary check, but can be very beneficial to researchers. An example would be providing information to help users create code to add labels to categorical variables.[10]

We hope that the proposed checks for data holders who plan to supply LFSD to potential users will be useful and will play a part in allowing administrative data to provide information for public benefit.

# 8 Acknowledgments

Part of Gillian Raab's time in 2024 was supported by Research Data Scotland.

# 9 Statement on conflicts of Interest

None that we are aware of.

# 10 Ethics Statement

Although some of the individual studies that have informed our practice have required ethical permission, one was required for this paper because no data from individuals or organisations was used in its preparation. We hope that readers will think that the proposals we have set forth in this paper will go some way to answer ethical concerns for LFSD.

# 11 References

[1] Sudlow C (2024) Uniting the UK's Health Data: A Huge Opportunity for Society. HDR UK, available from https://www.hdruk.ac.uk/helping-with-health-data/the-sudlow-review/

[2] Penner AM, Dodge KA. (2019) Using administrative data for social science and policy. RSF Russell Sage Found J Soc Sci. 2019;5(3):1–18. https://doi.org/10.7758/RSF.2019.5.3.01

[3] Nowok B., Raab G.M., Dibben C, (2017) Providing bespoke synthetic data for the UK Longitudinal Studies and other sensitive data with the synthpop package for R. *Statistical Journal of the IAOS,* 33(3):785-796; DOI: 10.3233/SJI-150153

[4] Dibben, C, Raab G.M., Nowok, B., Williamson, L., Adair, L. (2024) "Synthpop: A Tool to Enable More Flexible Use of Sensitive Data within the Scottish Longitudinal Study "in Drechsler, Jörg.(ed) Handbook of Sharing Confidential Data : Differential Privacy, Secure Multiparty Computation, and Synthetic Data. S.l: CHAPMAN and HALL CRC, 2024.4**.**

---

[10] The provision of JSON files is one option, see https://en.wikipedia.org/wiki/JSON

[5] Van Kesteren, E (2024) To democratize research with sensitive data, we should make synthetic data more accessible, Patterns, Volume 5, Issue 9, https://doi.org/10.1016/j.patter.2024.101049

[6] Rubin DB. (1993) Statistical disclosure limitation. J Off Stat. 1993;9(2):461–8.

[7] Reiter, J.: Synthetic data: A look back and a look forward. Transactions in Data Privacy 16, 15–-24

[8] Drechsler and Haensch 2023) Drechsler, J., Haensch, A.C.: 30 years of synthetic data. Statist. Sci. 39(2), 221‑242 (2024).

[9] Bulmer, M., & Coote, L. (2022). The role of synthetic data in teaching and learning statistics. In *Bridging the gap: empowering & educating today's learners in statistics. Proceedings of the 11th international conference on teaching statistics*. available from . https://iase-web.org/icots/11/proceedings/pdfs/ICOTS11_422_BULMER.pdf

[10] Kokosi T, De Stavola B, Mitra R, Frayling L, Doherty A, Dove I, Sonnenberg P, Harron K. An overview of synthetic administrative data for research. Int J Popul Data Sci. 2022 May 23;7(1):1727. doi: 10.23889/ijpds.v7i1.1727. PMID: 37650026; PMCID: PMC10464868.

[11] Office for National Statistics (2021). ONS methodology working paper series number 16 - Synthetic data pilot [Internet]. Available from: https://www.ons.gov.uk/methodology/methodologicalpublications/generalmethodology/onsworkingpaperseries/onsmethodologyworkingpaperseriesnumber16syntheticdatapilot

[12] Little, C., Elliot, M., and Allmendinger, R. (2022) Comparing the utility and disclosure risk of synthetic data with samples of microdata. In Privacy in Statistical Databases (Cham, 2022), J. Domingo-Ferrer and M. Laurent, Eds., Springer International Publishing, pp. 234-249.

[13] Raab, Gillian M, (2024) *Privacy Risk from Synthetic Data: Practical Proposals*. in Melek Önen, and Josep Domingo-Ferrer. Eds. *Privacy in Statistical Databases*. Cham: Springer Nature Switzerland, 2024. 254–273. Web.

[14] Calcraft P, Thomas I, Maglicic M, Sutherland A.(2021) ADRUK, Accelerating public policy research with synthetic data [Internet]. Available from: https://www.adruk.org/fileadmin/uploads/adruk/Documents/Accelerating_public_policy_research_with_synthetic_data_December_2021.pdf

[15] Office for National Statistics. (2023). *Annual Survey of Hours and Earnings, 2020: Synthetic Data Pilot*. [data collection]. UK Data Service. SN: 9045, DOI: http://doi.org/10.5255/UKDA-SN-9045-1

[16] ADR UK (2023) An interim ADR UK position statement on synthetic data https://www.adruk.org/fileadmin/uploads/adruk/Documents/An_interim_ADR_UK_position_statement_on_synthetic_data.pdf

[17] Research Data Scotland (2024), Synthetic census data now available for research. Web link =https://www.researchdata.scot/news-and-insights/synthetic-census-data-now-available-for-research/, accessed 22/12/024.

[18] Grath-Lone LM, Jay MA, Blackburn R, Gordon E, Zylbersztejn A, Wiljaars L, Gilbert R. What makes administrative data "research-ready"? A systematic review and thematic analysis of published literature. Int J Popul Data Sci. 2022 Apr 27;7(1):1718. doi: 10.23889/ijpds.v6i1.1718. PMID: 35520099; PMCID: PMC9052961

[19] Elliot, M., Mackey, E., and O'Hara, K. (2020) The anonymisation decision-making framework: European practitioners. Available from https://ukanon.net/framework/, 2020. Accessed: 2022-02-23.

[20] Hundepool, A., Domingo-Ferrer, J., Franconi, L., Giessing, S., Nordholt, E. S., Spicer, K., & De Wolf, P. P. (2012). *Statistical disclosure control*. John Wiley & Sons.

[21] Templ, M (2017). *Disclosure Control for Microdata : Methods and Applications in R*. 1st ed. 2017. Cham: Springer International Publishing.

[22] Nowok B. (2017) Recognising real people in synthetic microdata: risk mitigation and impact on utility. Paper presented at the Joint UNECE/Eurostat work session on statistical data confidentiality; Skopje, North Macedonia, 20-22 September 2017. Available at https://unece.org/statistics/events/SDC2017

[23] Raab, Gillian M, Beata Nowok, and Chris Dibben. "Practical Privacy Metrics for Synthetic Data." *arXiv.org* (2024): also available as a vignette in versions of the synthpop package. Form 1.9-0 onwards.